\documentclass{article}
\usepackage{amsmath,amssymb,amsthm,graphicx}

\newcommand{\dv}{\,\mathrm{div}\,}
\newcommand{\rot}{\mathrm{rot}\,}
\newcommand{\const}{\mathrm{const}}
\newcommand{\bu}{\mathbf{u}}
\newcommand{\bx}{\mathbf{x}}
\newcommand{\bm}{\mathbf{m}}
\newcommand{\bH}{\mathbf{H}}

\begin{document}
\large

\title{ Multidimensional fluid motions with planar waves}

\author{Sergey V. Golovin}
\date{}

\maketitle

\begin{center}
Lavrentyev Institute of Hydrodynamics, Novosibirsk 630090, Russia\\
sergey@hydro.nsc.ru
\end{center}

\begin{abstract}
In the classical one-dimensional solution of fluid dynamics
equations all unknown functions depend only on time $t$ and
Cartesian coordinate $x$. Although fluid spreads in all directions
(velocity vector has three components) the whole picture of motion
is relatively simple: trajectory of one fluid particle from plane
$x=\const$ completely determines motion of the whole plane. Basing
on the symmetry analysis of differential equations we propose
generalization of this solution allowing movements in different
directions of fluid particles belonging to plane $x=\const$. At
that, all functions but an angle determining the direction of
particle's motion depend on $t$ and $x$ only, whereas the angle
depends on all coordinates. In this solution the whole picture of
motion superposes from identical trajectories placed under different
angles in 3D space. Orientations of the trajectories are restricted
by a finite relation possessing functional arbitrariness. The
solution describes three-dimensional nonlinear processes and
singularities in infinitely conducting plasma, gas or incompressible
liquid.
\end{abstract}


\bigskip

AMS classification scheme numbers: 76W05, 76M60, 35C05, 35N10

\section*{Introduction}
The widely-used simplification of fluid dynamics equations is an
assumption of one-dimensionality of the flow. It is proposed that
all unknown functions depend only on two variables: time $t$ and
Cartesian coordinate $x$. Motion of fluid particles is allowed in
all directions, however most of interesting processes (waves of
compression and rarefaction, strong and weak discontinuities, etc.)
take place along one spatial axis $Ox$. Components of the velocity
vector, thermodynamical and all another unknown functions are
constant on the planes $x=\const$ and change from one plane to
another. This solution is often referred to as fluid motion with
planar waves. Being comparatively easy for an analytical analysis,
this simplification provides a great deal of information about
qualitative properties of fluid motions. However, the classical
one-dimensional solution can not describe three-dimensional
processes in fluid which in fact might be significant for the
correct description of the picture of the flow.

In present work we generalize the described classical
one-dimensional solution with planar waves. In our solution velocity
vector is decomposed into two components, one of which is parallel
and another one is orthogonal to $Ox$ axis. Absolute values
(lengthes) of the components and both thermodynamical functions
(density and pressure) are supposed to depend only on $t$ and $x$.
This part of solution coincide with the classical one. However, the
angle of rotation of velocity vector about $Ox$ axis is supposed to
depend on all independent variables $(t,x,y,z)$. Presence of this
function gives the desired generalization of the classical solution.

The proposed representation of the solution was advised by the
theory of symmetry analysis of differential equations
\cite{LVO,Olver}. Indeed, from the symmetry analysis point of view,
the classical one-dimensional solution is an invariant one of rank 2
with respect to the admissible group of shifts along $Oy$ and $Oz$
axis. Whereas the generalized solution is a partially invariant one
\cite{LVO} with respect to the full group of plain isometries
consisting of shifts along $Oy$ and $Oz$ axes and rotation about
$Ox$ axis.

Class of generalized solutions is happened to be a contansive one.
It is described by a closed system of PDEs with two independent
variables, which in the special case coincide with classical
equations for one-dimensional fluid motions. The angle as a function
of four independent variables is determined on solutions of the
invariant system from a finite (not differential) relation, which
has a functional arbitrariness. The finite relation allows clear
geometrical interpretation. This gives opportunity to construct a
desired type of fluid motion by choosing appropriate arbitrary
function in the functional relation.

Plasma flow governed by the solution possesses a remarkable
property. Fluid particles belonging to the same initial plane
$x=\const$ at some moment of time circumscribe the same trajectories
in 3D space and have identical magnetic field lines attached.
However, each trajectory and magnetic field line has its own
orientation, which depends on the position of the fluid particle in
the initial plane. The orientation is given by the finite relation
with functional arbitrariness. Thus, with the same shape of
trajectories and magnetic field lines one can construct infinitely
many pictures of fluid motions by varying admissibly directions of
particles spreading.

Intensively studied in recent scientific literature solution of
ideal compressible or incompressible fluid equations which is called
``singular vortex'' or ``Ovsyannikov vortex''
\cite{LVOSingVort,Chup1,Chup4, Chup3,Pavl,GolovinSingVortMHD,
GolovinSingVortMHDInvSubm, SingVortGeom} can be treated as the
analogous generalization of one-dimensional motion with spherical
waves. In this solution absolute values of the tangential and normal
to spheres $r=\const$ components of velocity vector field depend
only on time $t$ and distance $r$ to the origin. An angle of
rotation of the vector field about the radial direction $Or$ is a
function on all independent variables. This solution also allows
symmetry interpretation as the partially invariant one with respect
to the admissible group of sphere isometries, i.e. of rotations in
$\mathbb{R}^3$.

The generalized one-dimensional solution with planar waves for ideal
gas dynamics equations was first obtained in \cite{LVOPIS}. For all
we known, it was not analyzed in details for its physical content.
In present work we observe equations of ideal magnetohydrodynamics.
Cases of ideal gas dynamics and ideal liquid can be obtained in
limits of zero magnetic field $\bH\equiv0$ and constant density
$\rho=\const$ respectively.

The paper is organized as follows. We start from the formulation of
the representation of solution, which is prescribed by symmetry
properties of the main model of ideal magnetohydrodynamics.
Substitution of the representation of the solution into the system
of equations brings a highly-overdetermined system of PDEs for the
non-invariant function
--- angle of rotation of the vector fields about $Ox$ axis.
Investigation of the overdetermined system reveals two main cases,
when some auxiliary function $h$ is equal or not equal to zero. From
the mechanical point of view these two cases correspond to the
compressible or incompressible (divergence-free) vector field which
is obtained as a projection of the velocity field into $Oyz$ plane.
In both cases the overdetermined system is reduced to some
compatible invariant subsystem of PDEs with two independent
variables and a finite implicit relation for the non-invariant
function. We give geometrical interpretation of the finite relation,
which allows keeping track of the singularities, which may take
place in the flow. We prove that particles trajectories and magnetic
field lines are planar curves. Moreover, these curves are the same
for all particles, which start from the same initial plane
$x=\const$. This gives opportunity to construct a pattern of the
trajectory and magnetic field line. The complete 3D picture of the
flow is obtained by attaching the pattern to every point in fixed
$Oyz$ plane in accordance to the directional field defined by the
finite relation for the non-invariant function. Remarkable, that the
same pattern of magnetic line and trajectory attached to different
directional field in $Oyz$ plane produces variety of pictures of
plasma motion in 3D space. As an example, the solution is used for
explicit description of the plasma flow in axisymmetric canal with
curved conducting walls.

\section{Representation of the solution and preliminary
analysis}\label{s1}
\subsection{Representation of the solution}

The system of ideal magnetohydrodynamics (tension comes to pressure,
thermal conductivity is zero, electric conductivity is infinite) has
the form \cite{KulikLubim}
\begin{eqnarray}\label{MHD-cont}
&&D\,\rho+\rho\dv\bu=0,\\\label{MHD-moment} &&D\,\bu+\rho^{-1}\nabla
p+\rho^{-1}\bH\times\rot\bH=0,\\\label{MHD-pressure}
&&D\,p+A(p,\rho)\dv\bu=0,\\\label{MHD-induction}
&&D\,\bH+\bH\dv\bu-(\bH\cdot\nabla)\bu=0,\\\label{MHD-Faradey}
&&\dv\bH=0,\;\;\;D=\partial_t+\bu\cdot\nabla.
\end{eqnarray}
Here $\bu=(u,v,w)$ is the fluid velocity vector, $\bH=(H,K,L)$ is
the magnetic vector field; $p$ and $\rho$ are  pressure and density.
The state equation $p=F(S,\rho)$ with the entropy $S$ gives rise to
function $A(p,\rho)$ defined by $A=\rho\,(\partial F/\partial\rho)$.
All unknown functions depend on time $t$ and Cartesian coordinates
$\bx=(x,y,z)$.

In the case of arbitrary state equation $p=F(S,\rho)$ equations
(\ref{MHD-cont})--(\ref{MHD-Faradey}) admit 11-dimensional Lie group
$G_{11}$ of point transformations, which is 10-dimensional Galilean
group extended by the homothety \cite{Fuchs,Ibragimov}. Optimal
system of subgroups $\Theta G_{11}$ was constructed in
\cite{LVOSubm,GrundLala}, see also \cite{LVOGDE01}. Examination of
$\Theta G_{11}$ shows, that the partially invariant solution of
described type is generated by 3-dimensional subgroup
$G_{3.13}\subset G_{11}$ with Lie algebra $L_{3.13}$ spanned by the
infinitesimal generators
$\{\partial_y,\,\partial_z,\,z\partial_y-y\partial_z+
w\partial_v-v\partial_w+L\partial_K-K\partial_L\}$ (we use the
subgroups numeration according to \cite{LVOGDE01}).

\begin{figure}
  % Requires \usepackage{graphicx}
  \includegraphics[width=0.4\textwidth]{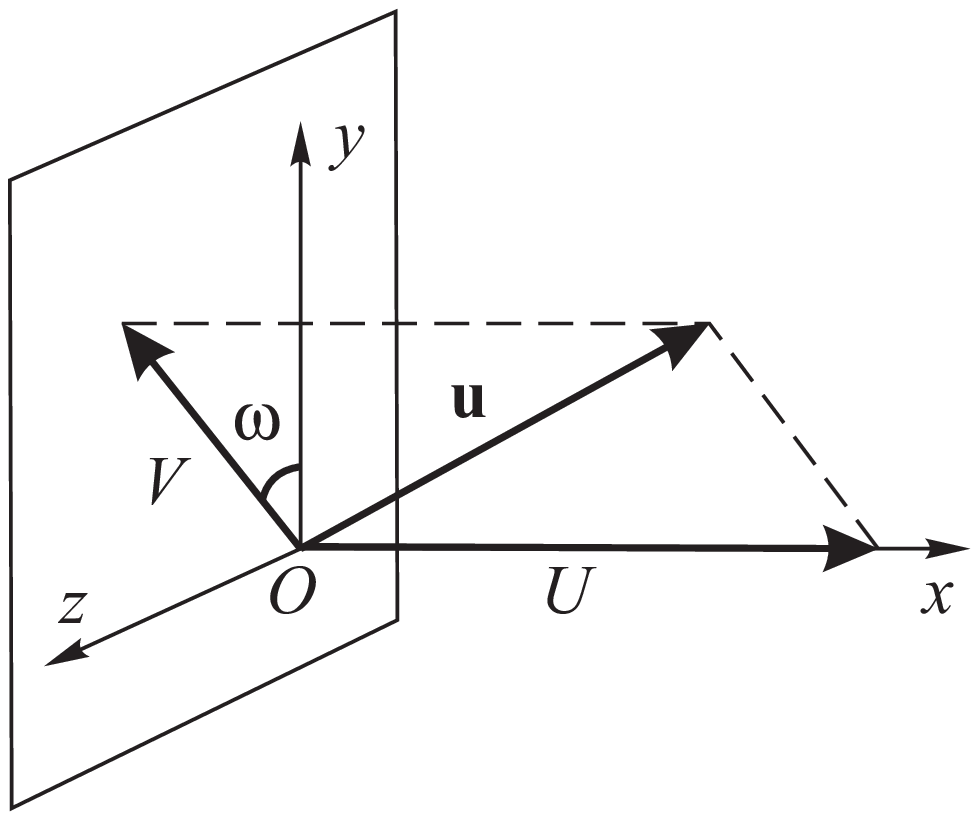}\hfill
  \includegraphics[width=0.4\textwidth]{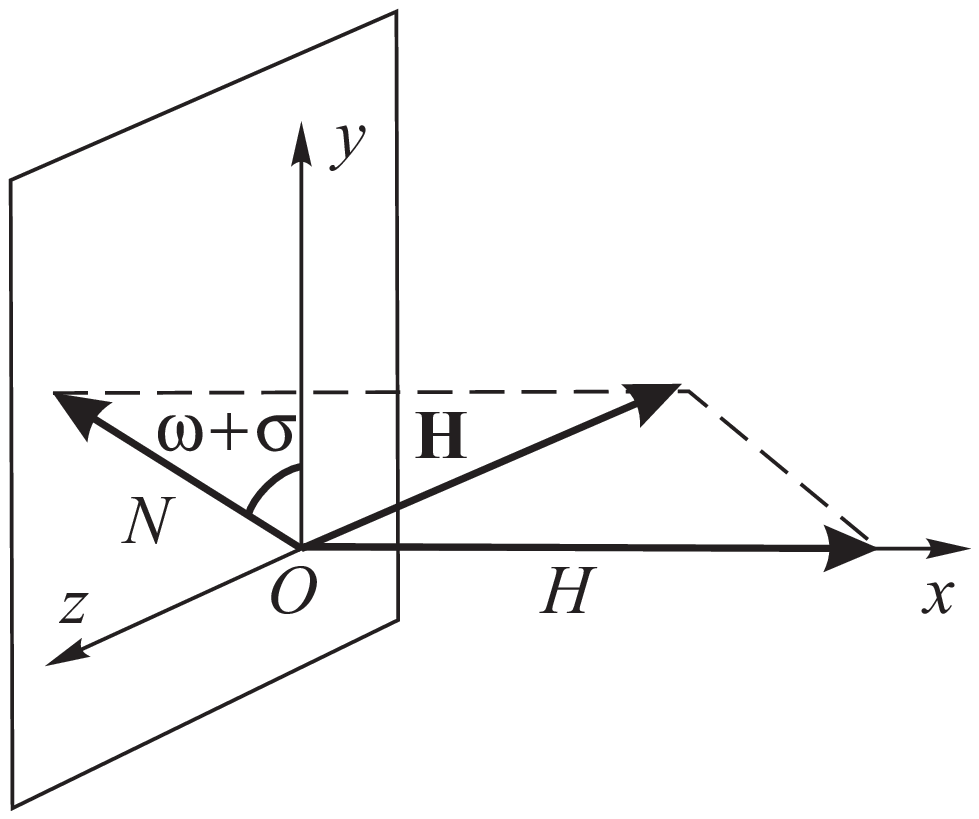}
  \caption{Representation of velocity vector $\bu$ and magnetic
  field vector $\bH$ in the partially invariant solution. All functions
  but $\omega$ depend on $t$ and
  $x$, whereas $\omega=\omega(t,x,y,z)$.}\label{coords}
\end{figure}

Indeed, Lie group $G_{3.13}$ is spanned by shifts along $Oy$ and
$Oz$ axes and simultaneous rotations about the first coordinate axis
in $\mathbb{R}^3(\bx)$, $\mathbb{R}^3(\bu)$, and
$\mathbb{R}^3(\bH)$. Invariants of this group of transformations in
the space of independent variables and dependent functions
$\mathbb{R}^4(t,\bx)\times\mathbb{R}^8(\bu,\bH,p,\rho)$ are
\begin{equation}\label{Invars}
t,\;\;\;x,\;\;\;u,\;\;\;V=\sqrt{v^2+w^2},\;\;\;p,\;\;\;
\rho,\;\;\;H,\;\;\;N=\sqrt{K^2+L^2},\;\;\mbox{ and }\;\;vK+wL.
\end{equation}
The last invariant may be treated as angle $\sigma$ between the
projections of vectors $\bu$ and $\bH$ into $Oyz$ plane (see figure
\ref{coords}). The general theory of partially invariant solutions
may be found in \cite{LVO}. The representation of partially
invariant solution is obtained by assigning a functional dependence
between the group invariants (\ref{Invars}). In particular, for the
solution of rank 2 (two invariant independent variables) and defect
1 (one non-invariant function) it gives the following representation
of solution:
\begin{equation}\label{SolRepr}
\begin{array}{l}
\begin{array}{ll}
u=U(t,x),&H=H(t,x),\\[2mm]
v=V(t,x)\cos\omega(t,x,y,z),\;\;&K=N(t,x)\cos\big(\omega(t,x,y,z)+\sigma(t,x)\big),\\[2mm]
w=V(t,x)\sin\omega(t,x,y,z),&L=N(t,x)\sin\big(\omega(t,x,y,z)+\sigma(t,x)\big),\\[2mm]
\end{array}\\
\;\;p=p(t,x),\;\;\;\rho=\rho(t,x),\;\;\;S=S(t,x).
\end{array}
\end{equation}
Here only the non-invariant function $\omega(t,x,y,z)$ depends on
all original independent variables. Functions $U$, $V$, $H$, $N$,
$\sigma$, $p$, $\rho$ are invariant with respect to $G_{3.13}$. They
depend only on invariant variables $t$ and $x$. These functions will
be referred to as invariant ones. The system of equations for
determination of invariant and non-invariant functions will be
called the submodel of the main model of ideal magnetohydrodynamics.

\subsection{Analysis of the submodel} Substitution of the
representation (\ref{SolRepr}) into
(\ref{MHD-cont})--(\ref{MHD-Faradey}) gives the following result.
The continuity equation (\ref{MHD-cont}) allows introduction of new
unknown invariant function $h(t,x)$, defined by the following
relation
\begin{equation}\label{Inv1}
\widetilde{D}\,\rho+\rho(U_x+hV)=0.
\end{equation}
Hereinafter $\widetilde{D}$ denotes the invariant part of the
differentiation along the trajectory
\[\widetilde{D}=\partial_t+U\partial_x.\]
The remaining part of the continuity equation gives an equation for
function $\omega$:
\begin{equation}\label{NonInv1}
\sin\omega\,\omega_y-\cos\omega\,\omega_z+h=0.
\end{equation}
Another equations for invariant functions follow from the first
components of momentum (\ref{MHD-moment}) and induction equations
(\ref{MHD-induction}), and also pressure equation
(\ref{MHD-pressure}).
\begin{eqnarray}\label{Inv2}
&&\widetilde{D}\,U+\rho^{-1}p_x+\rho^{-1}NN_x=0,\\\label{Inv3}
&&\widetilde{D}\,H+hHV=0,\\\label{Inv41}
&&\widetilde{D}\,p+A(p,\rho)(U_x+hV)=0.
\end{eqnarray}
The rest of system (\ref{MHD-cont})--(\ref{MHD-Faradey}) gives rise
to the overdetermined system for function $\omega$. From a
nondegenerate linear combination of equations (\ref{MHD-moment}) in
projections to $Oy$ and $Oz$ axes one obtains
\begin{eqnarray}\label{NonInv2}
\rho V\omega_t+\big(\rho \,UV-HN\cos\sigma\big)\,\omega_x+\big(\rho
V^2\cos\omega-N^2\cos\sigma\cos(\omega+\sigma)\big)\,\omega_y
\\[2mm]\nonumber
+ \big(\rho
V^2\sin\omega-N^2\cos\sigma\sin(\omega+\sigma)\big)\,\omega_z
-H(N_x\sin\sigma+N\cos\sigma\sigma_x)=0.
\end{eqnarray}
\begin{eqnarray}\label{NonInv3}
HN\sin\sigma\,\omega_x+N^2\sin\sigma\cos(\omega+\sigma)\,\omega_y+
N^2\sin\sigma\sin(\omega+\sigma)\,\omega_z\\[2mm]\nonumber
+\rho\,\widetilde{D}V+HN\sin\sigma\,\sigma_x-HN_x\cos\sigma=0.
\end{eqnarray}
The same operation with remaining two induction equations
(\ref{MHD-induction}) provides
\begin{eqnarray}\label{NonInv4}
N\omega_t+(NU-HV\cos\sigma)\,\omega_x+VN\sin\sigma\sin(\omega+\sigma)\,\omega_y\\[2mm]\nonumber
-VN\sin\sigma\cos(\omega+\sigma)\,\omega_z+N\widetilde{D}\sigma+HV_x\sin\sigma=0.
\end{eqnarray}
\begin{eqnarray}\label{NonInv5}
HV\sin\sigma\,\omega_x+NV\cos\sigma\sin(\omega+\sigma)\,\omega_y\\[2mm]\nonumber
-NV\cos\sigma\cos(\omega+\sigma)\,\omega_z-\widetilde{D}N+HV_x\cos\sigma-NU_x=0.
\end{eqnarray}
Finally, equation (\ref{MHD-Faradey}) is transformed to
\begin{equation}\label{NonInv6}
N\big(\sin(\omega+\sigma)\omega_y-\cos(\omega+\sigma)\,\omega_z\big)-H_x=0.
\end{equation}

The overdetermined system (\ref{NonInv1}),
(\ref{NonInv2})--(\ref{NonInv6}) for non-invariant function $\omega$
should be investigated for compatibility \cite{Pommaret}. At that we
observe only solution with functional arbitrariness in determination
of function $\omega$. This condition, in particular, implies
non-reducibility of the solution to the classical one-dimensional
solution with planar waves.

Function $\omega$ determines with only constant arbitrariness if it
is possible to express all first-order derivatives of $\omega$ from
the system of equations (\ref{NonInv1}),
(\ref{NonInv2})--(\ref{NonInv6}). In order to prohibit this
situation one should calculate a matrix of coefficients of the
derivatives of function $\omega$ and vanish all its rank minors.
This leads to the following four cases:
\begin{equation}\label{Cases}
1.\;H=0;\;\;\;2.\;N=0;\;\;\;3.\;V=0;\;\;\;\;4.\;\sigma=0\, \mbox{ or
}\sigma=\pi.
\end{equation}
By definition (\ref{SolRepr}) functions $V$ and $N$ are
non-negative. Values $\sigma=\pi$ and $\sigma=0$ in the case 4
(\ref{Cases}) differ only by the sign of function $N$. Both can be
observed in the same framework for $\sigma=0$, non-negative $V$ and
arbitrary $N$.

Cases 2 and 3 in classification (\ref{Cases}) correspond to the
magnetic field or velocity parallel to $Ox$-axis. Both of them embed
into the case $\sigma=0$. Indeed, if $\sigma=0$ then the velocity
vector at each particle and its magnetic field vector belong to the
plane, which is orthogonal to $Oyz$ coordinate plane. Therefore,
cases 2 and 3 are degenerate versions of this more general
situation. Case 4 will be observed as the main case in the following
calculations. In case of pure gas dynamics $\bH\equiv0$ three of
four conditions (\ref{Cases}) satisfied automatically, hence the
solution is irreducible without any additional restrictions.

\subsection{Case of planar magnetic field} Let
us first observe the case $H=0$, when the magnetic field vector is
parallel to $Oyz$ plane. The compatibility condition of equations
(\ref{NonInv1}) and (\ref{NonInv6}) in this case is
\begin{equation}\label{Compat1}
\big(\cos(\omega+\sigma)\,\omega_y+\sin(\omega+\sigma)\,\omega_z\big)h=0.
\end{equation}
For $h=0$ the determinant of the homogenous system of algebraic
equations (\ref{NonInv1}), (\ref{NonInv6}) for $\omega_y$ and
$\omega_z$ is $\sin\sigma$. Hence, the solution is non-trivial only
for  $\sigma=0$ or $\sigma=\pi$. The case $h\ne0$ leads to the
reduction following from equations (\ref{NonInv6}) and
(\ref{Compat1}). Thus, the non-trivial solution exists only for
$\sin\sigma=0$, i.e. case 1 in the classification (\ref{Cases})
contains in case 4.

\section{The main case \boldmath$h\ne0$ }\label{s2}
\subsection{Equations of the submodel} Let us observe the main case $\sigma=0$.
From the mechanical point of view it corresponds to a plasma flow
where velocity and magnetic field vectors at each particle are
coplanar to $Ox$ axis. Equations (\ref{Inv1}),
(\ref{Inv2})--(\ref{Inv4}) belong to the invariant part of the
submodel. Besides, equation (\ref{NonInv3}) gives
\begin{equation}\label{Inv31}
\widetilde{D}\,V-\rho^{-1}HN_x=0.
\end{equation}
From equation (\ref{NonInv5}) taking into account (\ref{NonInv1})
one obtains
\begin{equation}\label{Inv32}
\widetilde{D}\,N+NU_x-HV_x+hNV=0.
\end{equation}
Finally, equation (\ref{NonInv6}) due to the relation
(\ref{NonInv1}) can be written as
\begin{equation}\label{Inv33}
H_x+hN=0.
\end{equation}
In addition to the equation (\ref{NonInv1}), the non-invariant part
of the determining system contains two equations, which follow from
(\ref{NonInv2}), (\ref{NonInv4}):
\begin{eqnarray}\label{NonInv31}
&&\rho V\omega_t+\big(\rho \,UV-HN\big)\,\omega_x+\big(\rho
V^2-N^2\big)(\cos\omega\,\omega_y
+\sin\omega\,\omega_z)=0,\\\label{NonInv32}
&&N\omega_t+(NU-HV)\,\omega_x=0.
\end{eqnarray}
Elimination of the derivative $\omega_t$ from equations
(\ref{NonInv31}), (\ref{NonInv32}) gives the following classifying
relation
\begin{equation}\label{Class3}
(\rho V^2-N^2)\big(H\omega_x+N(\cos
\omega\,\omega_y+\sin\omega\,\omega_z)\big)=0.
\end{equation}
We observe only the case when the second factor in (\ref{Class3})
vanishes. The compatibility conditions of equations (\ref{NonInv1}),
(\ref{NonInv32}), and (\ref{Class3}) are
\begin{eqnarray}\label{Inv34}
&&N\widetilde{D}\,h-HVh_x=0,\\\label{Inv35} &&Hh_x+h^2N=0.
\end{eqnarray}
For $h\ne0$ there is an integral
\begin{equation}\label{Integ31}
H=H_0h,\;\;\;H_0=\const.
\end{equation}
Thus, the submodel's equations are reduced to the following ones.
\begin{eqnarray}\label{main1}
&&\widetilde{D}\,\rho+\rho(U_x+hV)=0.\\\label{main2}
&&\widetilde{D}\,U+\rho^{-1}p_x+\rho^{-1}NN_x=0.\\\label{main3}
&&\widetilde{D}\,V-\rho^{-1}H_0hN_x=0,\\\label{main4}
&&\widetilde{D}\,p+A(p,\rho)(U_x+hV)=0,\\\label{main5}
&&\widetilde{D}\,N+NU_x-H_0hV_x+hNV=0,\\\label{main6}
&&\widetilde{D}\,h+Vh^2=0,\;\;\;H_0h_x+hN=0.
\end{eqnarray}
The obtained system \eqref{main1}--\eqref{main6} inherits the
overdetermination of the initial MHD equations
\eqref{MHD-cont}--\eqref{MHD-Faradey}. However, its compatibility
conditions satisfied by virtue of the system itself. Indeed, the
only nontrivial compatibility condition of the system
\eqref{main1}--\eqref{main6} is given by two equations \eqref{main6}
for function $h$. Cross-differentiation of \eqref{main6} shows that
their compatibility condition coincide with equation \eqref{main5},
i.e. is already contained in the system. The most general Cauchy
problem for system \eqref{main1}--\eqref{main6} requires assigning
functions $\rho$, $U$, $V$, $p$, $N$ at $t=0$ as functions of $x$,
and fixing a constant value of $h$ at $t=0$ over some plane
$x=\const$. For pure gas dynamics $\bH\equiv0$ the second equation
\eqref{main6} satisfies identically, hence the initial data for $h$
become $h(0,x)=h_0(x)$. System \eqref{main1}--\eqref{main6} equipped
by the suitable initial data can be solved numerically. It also
allows exact reductions to systems of ODEs since the admitted
symmetry group is obviously nontrivial.

Equations \eqref{NonInv31}--\eqref{Class3} for the non-invariant
function can be integrated. Function $\omega$ determines by the
following implicit equation
\begin{equation}\label{omega3}
F(y-\tau\cos\omega,\;z-\tau\sin\omega)=0
\end{equation}
with $\tau=1/h$ and arbitrary smooth function $F$. In case of pure
gas dynamics $\bH\equiv 0$ equation (\ref{NonInv32}) identically
satisfied. Therefore, function $F$ in the general formula
(\ref{omega3}) for function $\omega$ also arbitrarily depends on
$\xi$: $\xi_t+U\xi_x=0$. Results of the performed calculations are
summarized in the following theorem.

{\bf Theorem \ref{s2}.} {\it In the main case $\sigma=0$ and $h\ne0$
the invariant functions are determined by the system of differential
equations (\ref{main1})--(\ref{main6}). The non-invariant function
$\omega$ is given by the implicit equation (\ref{omega3}) with
arbitrary smooth function $F$.}

\subsection{Geometrical construction of the field of
directions}\label{smain}

\begin{figure}[t]
  \centering
  \includegraphics[width=0.4\textwidth]{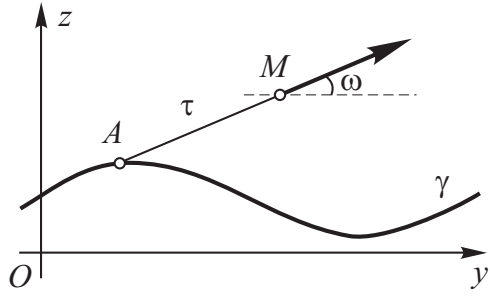}\\
  \caption{Geometric interpretation of the solution $\omega=\omega\bigl(\tau(t,x),y,z\bigr)$ of the
  implicit equation (\ref{omega3}). Curve $\gamma:\,F(y,z)=0$ is determined by the same
  function $F$ as in (\ref{omega3}). Function $\omega$ at given point $M$ is the angle between
  the direction of line segment $AM$ and $Oy$ axis, where $A\in\gamma$ and $|AM|=\tau$.}\label{f1}
\end{figure}

\begin{figure}[t]
  \centering
  \includegraphics[width=0.6\textwidth]{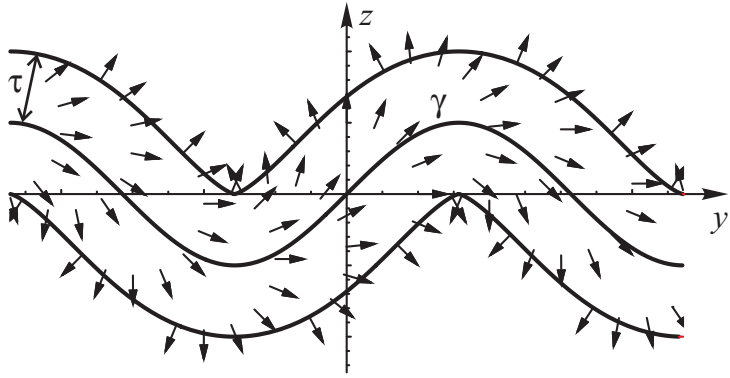}\\
  \caption{The field of directions is defined by the implicit equation (\ref{omega3})
  in the stripe of determinacy of width $2\tau$ with curve $\gamma:\,F(y,z)=0$ as a medial
  line. In this example $F=z-\sin y$. At the points of limiting equidistants
  the field of directions is orthogonal to the equidistants.
  }\label{VectMain}
\end{figure}

Here we give an algorithm for solving the implicit relation
(\ref{omega3}) over some fixed plane $x=x_0$ at time $t=t_0$.
Suppose that function $F$ in (\ref{omega3}) is fixed. This specifies
a curve $\gamma=\{(y,z)\,|\,F(y,z)=0\}$. In order to find angle
$\omega$ at arbitrary point $M=(y,z)$ one should draw a line segment
$AM$ of the length $\tau$ such that $A\in\gamma$. The direction of
$AM$ gives the required angle $\omega$ as it is shown in figure
\ref{f1}. Function $\omega$ is only defined at points located within
distance $\tau$ from the curve $\gamma$. The rest of $Oyz$ plane
does not belong to the domain of $\omega$. Boundaries of the domain
of $\omega$ are $\tau$-equidistants to $\gamma$. As $x$ grows,
function $\tau$ changes according to the solution of equations
\eqref{main1}--\eqref{main6}. This prescribes modification of the
$\omega$-domain over different planes $x=\const$. Thus, the domain
of function $\omega$ (hence, of the whole solution (\ref{SolRepr}))
over each plane $x=\const$ is a stripe of determinacy of the width
$2\tau$ with curve $\gamma$ as a centerline (see figure
\ref{VectMain}). The stripe of determinacy is bounded by
equidistants curves to $\gamma$. Over the boundaries of
$\omega$-domain the field of directions $\omega$ is orthogonal to
the boundaries.

Inside its domain function $\omega$ is multiply-defined. Indeed,
there are could be several line segments $AM$ with $A\in\gamma$
giving rise to several branches of function $\omega$. However, it is
always possible to choose a single-valued and continuous branch of
$\omega$.

\begin{figure}[t]
\centering
\includegraphics[width=0.32\textwidth]{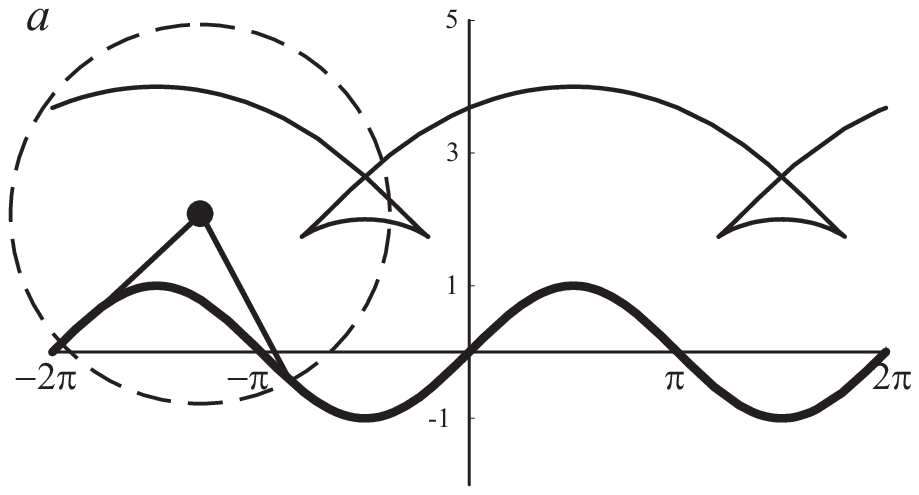}\hfill
\includegraphics[width=0.32\textwidth]{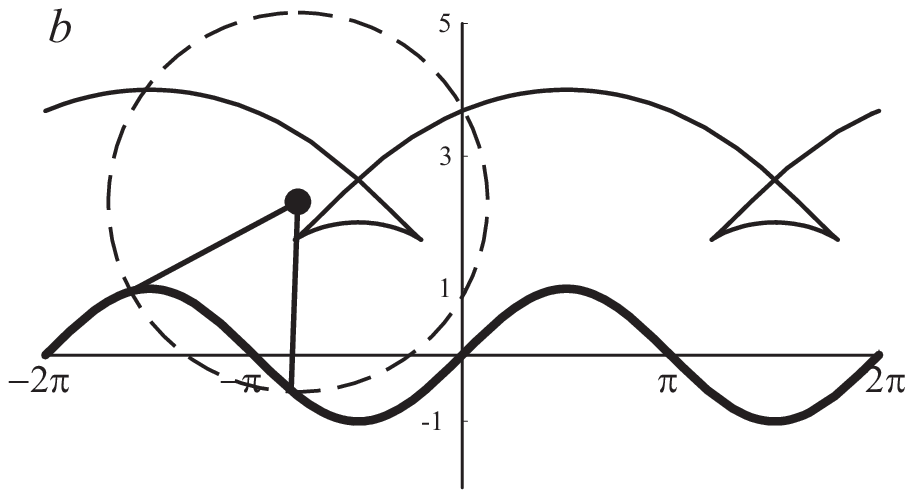}\hfill
\includegraphics[width=0.32\textwidth]{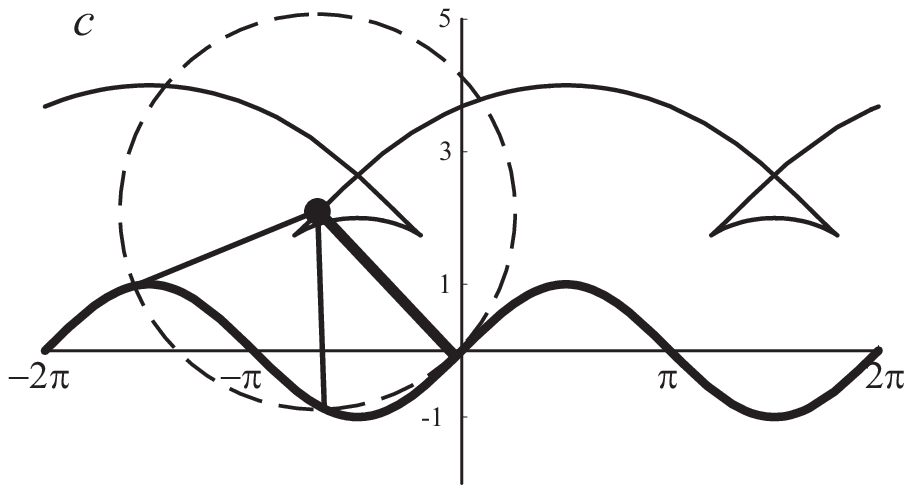}\\[2mm]
\includegraphics[width=0.32\textwidth]{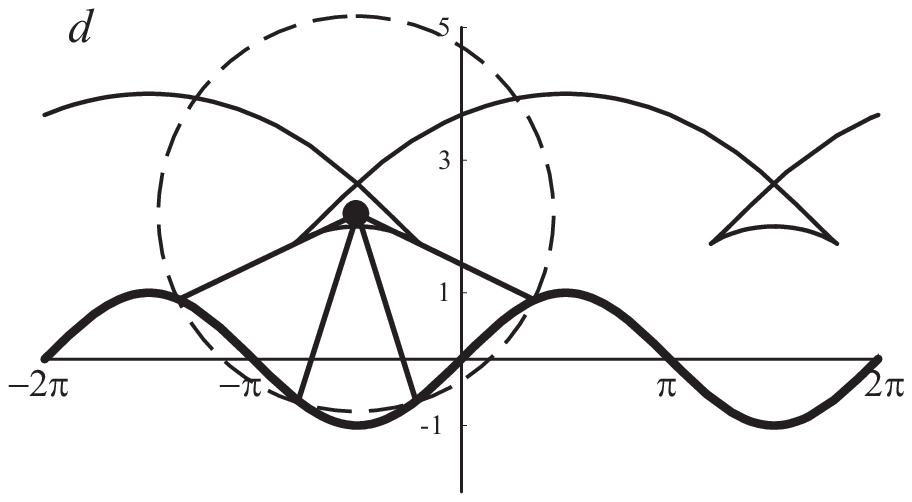}\hfill
\includegraphics[width=0.32\textwidth]{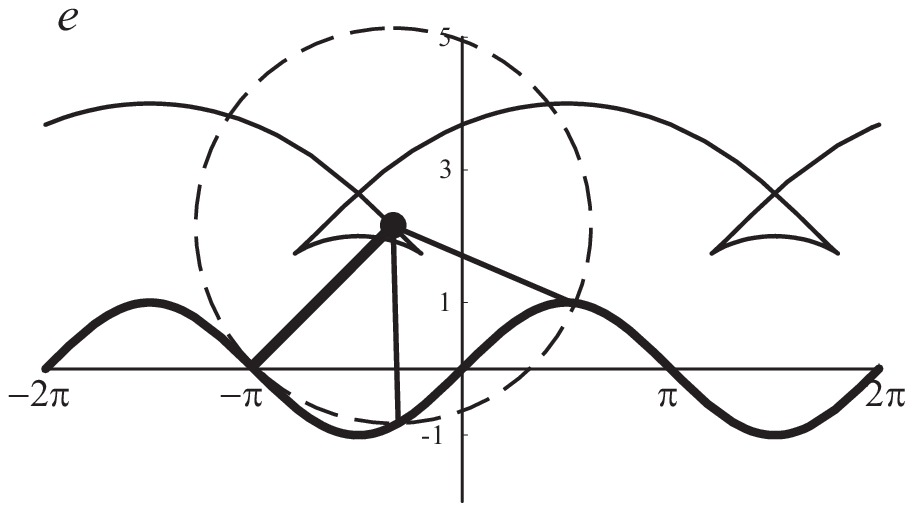}\hfill
\includegraphics[width=0.32\textwidth]{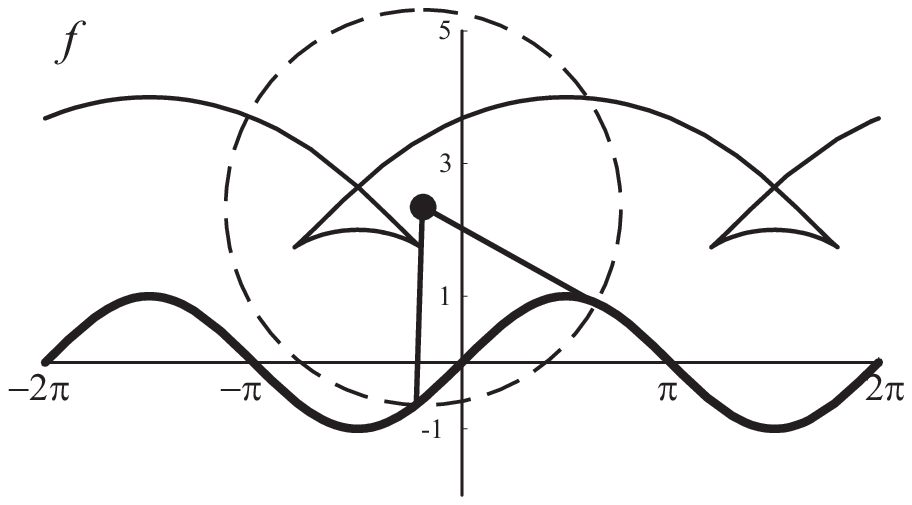}
\caption{The behaviour of function $\omega$ over the
dovetail. There are two branches of $\omega$ outside the dovetail in
Figures (a), (b) and (f); three branches of $\omega$ at the borders
of the dovetail in Figures (c) and (e); and four branches of
solution inside the dovetail in Figure (d).}\label{f5}\label{dovetail}
\end{figure}

Discontinuities of $\omega$ may appear in cases when the
equidistants to $\gamma$ have the dovetail singularities. The
observations illustrated by figure \ref{dovetail} show that every
branch of function $\omega$ necessary have a line of discontinuity
inside or at the border of the dovetail. In figure 4 the curve
$\gamma$ is a sinusoid shown at the bottom of figures; the curve on
the top is the equidistant shifted at large enough distance $\tau$.
For the convenience we draw the circle of radius $\tau$ with center
at chosen point $M$. Each intersection of the circle with $\gamma$
gives rise to a branch of $\omega$. Let us take $M$ outside of the
dovetail (figure {\it a}). There are two branches of $\omega$ at
$M$. As $M$ moves towards the borders of the dovetail, both branches
change continuously (figure {\it b}). At the border of the dovetail
the new branch of $\omega$ appears (figure {\it c}). The latter
splits into two branches inside the dovetail (figure {\it d}). As
$M$ reaches the right boundary of the dovetail the two "old"
branches of $\omega$ sticks together (figure {\it e}) and disappear
as $M$ leaves the dovetail (figure {\it f}). One can chase, that the
branches of $\omega$ obtained on the right-hand side of the dovetail
are different from the ones existed on the left-hand side of the
dovetail.

The dovetails do not appear if $\tau<\min\limits_{\bx\in\gamma}
R(\bx)$, where $R(\bx)$ is a curvature radius of curve $\gamma$ at
$\bx$. So, one can avoid the singularities either by choosing the
solution with small enough $\tau$ or by fixing the curve $\gamma$
with large curvature radius. Described discontinuities takes the
solution out of class \eqref{SolRepr}. They can not be interpreted
in shock waves framework. Indeed, over the line of discontinuity
only the direction of the magnetic and velocity vector fields
change, while their absolute values together with thermodynamics
functions remain continuous. Another type of transverse or
alfv\'{e}ic waves \cite{KulikLubim, JeffreyTaniuti} characteristic
to ideal MHD equations also can not explain the discontinuity as
long as the magnetic and velocity fields rotates not across the
front of discontinuity.

Appearance of the dovetail singularities physically mean magnetic
field lines, which pass through different point in some initial
plane $x=\const$ collide in their further development. This happens
if the function $\tau$ increases along the magnetic lines such that
the $\tau$-equidistants to $\gamma$ became non-smooth. In the
vicinity of the collision point the solution leaves the prescribed
class \eqref{SolRepr}; the corresponding fluid flow should be
observed either in general 3D framework, or in terms of an extended
main model, i.e. taking into account magnetic or kinematic viscosity
as it is observed in magnetic reconnection problems
\cite{PriestForbes}. This nonlinear process is specific to the
constructed solution, and can not take place in the classical
one-dimensional solution with planar waves, where all magnetic lines
are parallel to each other.

\section{\boldmath Case $h=0$}\label{s3}
\subsection{Equations of the submodel}
From the mechanical point of view this case means that the
projection of vector field $\bu$ into the plane $x=\const$ is
incompressible, i.e. its divergence is zero. This case is observed
separately because the non-invariant function $\omega$ is determined
by different algorithm.

For $h=0$ integral (\ref{Integ31}) is not valid. Instead, equations
(\ref{Inv3}) and (\ref{Inv33}) give
\[H=H_0=\const.\]
Thus, equations of the invariant system are
\begin{equation}\label{Inv4}
\begin{array}{l}
\widetilde{D}\,\rho+\rho\,U_x=0,\\[2mm]
\widetilde{D}\,U+\rho^{-1}p_x+\rho^{-1}NN_x=0,\\[2mm]
\widetilde{D}\,V-\rho^{-1}H_0N_x=0,\\[2mm]
\widetilde{D}\,p+A(p,\rho)\,U_x=0,\\[2mm]
\widetilde{D}\,N+NU_x-H_0V_x=0.
\end{array}
\end{equation}
This system of 5 equations serves for determination of 5 unknown
functions $U$, $V$, $N$, $p$, and $\rho$. The non-invariant function
$\omega$ is restricted by equations (\ref{NonInv1}),
(\ref{NonInv32}), and (\ref{Class3}). Suppose that  its solution
$\omega=\omega(t,x,y,z)$ for $N\ne0$ and $\rho V^2-N^2\ne0$ is given
implicitly by the equation $\Phi(t,x,y,z,\omega)=0$,
$\Phi_\omega\ne0$. The system (\ref{NonInv1}), (\ref{NonInv32}), and
(\ref{Class3}) transforms as follows
\begin{equation}\label{NonInvh0}
\Phi_k=0,\;\;\;\Phi_t+U\Phi_x+V\Phi_j=0,\;\;\;H_0\,\Phi_x+N\Phi_j=0.
\end{equation}
Here $Ojk$ is a Cartesian frame of reference rotated on angle
$\omega$ about the origin.
\begin{equation}\label{Ojk}
j=y\cos \omega+z\sin\omega,\;\;\;k=-y\sin\omega+z\cos\omega.
\end{equation}
Integrals of system (\ref{NonInvh0}) are $\omega$ and
$j-\varphi(t,x)$, where function $\varphi(t,x)$ satisfies the
overdetermined system
\begin{equation}\label{Eqvarphi}
\varphi_t+U\varphi_x=V,\;\;\;H_0\,\varphi_x=N.
\end{equation}
The compatibility condition of equations (\ref{Eqvarphi}) is the
last equation of the invariant system (\ref{Inv4}). Differential
one-form
\[H_0d\varphi=(H_0V-NU)dt+Ndx\]
is closed, therefore function $\varphi$ can be found by integration
as
\[\varphi(t,x)=\int\limits_{(t_0,x_0)}^{(t,x)}d\varphi.\]
Note, that the initial data for function $\varphi$ is given by only
one constant $\varphi(t_0,x_0)$. The non-invariant function $\omega$
can be taken in the form of the finite implicit equation
\begin{equation}\label{ImplEqh0}
j=f(\omega)+\varphi(t,x)
\end{equation}
with arbitrary smooth function $f$. The result is formulated in the
following theorem.

{\bf Theorem \ref{s3}.} {\it In the case $\sigma=h=0$ the invariant
functions are determined from equations (\ref{Inv4}),
(\ref{Eqvarphi}). Function $\omega$ is given by the implicit
equation (\ref{ImplEqh0}).}

\subsection{Construction and properties of the field of directions}

\begin{figure}[t]
\begin{minipage}[b]{0.5\textwidth}
\centering
\includegraphics[width=0.9\textwidth]{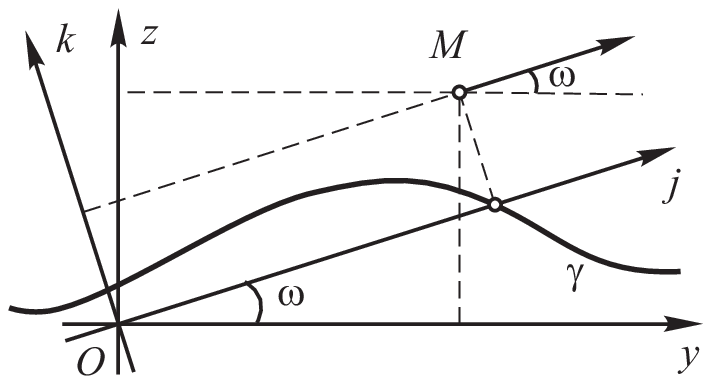}
\caption{Given a value of $\omega$ at some point $M$, the auxiliary
$Ojk$ frame of reference is defined as shown. The projection of $M$
into the $Oj$ axis is called the base point for $M$. The set of all
the base points for different $M$ with different $\omega(M)$ forms
the basic curve $\gamma$.}\label{fh_0_1}
\end{minipage}
\hspace{3mm}
\begin{minipage}[b]{0.5\textwidth}
\centering
\includegraphics[width=0.9\textwidth]{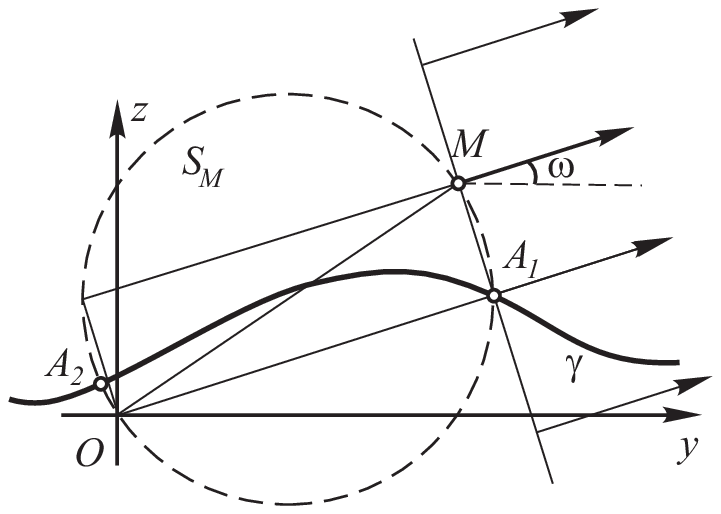}
\caption{Given curve $\gamma$ one can find $\omega$ at any point $M$
of the corresponding $Oyz$ plane. Circle $S_M$ with diameter $OM$
should be drawn. Let $A_i$ be points of intersection of $S_M$ with
$\gamma$. For each $A_i$ the angle $\omega$ at $M$ is given by the
direction $OA_i$ as shown. }\label{fh_0_2}
\end{minipage}
\end{figure}

Now we clarify a geometrical interpretation of the implicit relation
(\ref{ImplEqh0}). Let us fix a plane $x=x_0$ and time $t=t_0$. For
simplicity we assume $\varphi(t_0,x_0)=0$. Let the value of $\omega$
satisfying (\ref{ImplEqh0}) is known at some point $M=(y,z)$ of the
plane $x=x_0$. Consider a Cartesian frame of reference $Ojk$ turned
counterclockwise on angle $\omega$ with respect to $Oyz$ (see figure
\ref{fh_0_1}). By the construction, $j$-coordinate of point $M$ and
angle $\omega$ are related by $j=f(\omega)$. All points with the
same coordinate $j$ and arbitrary coordinate $k$ satisfy the same
relation.
A point satisfying the relation (\ref{ImplEqh0}) with zero
coordinate $k$ will be referred to as the base point for chosen
values of $j$ and $\omega$. The locus of all base points for various
$j$ and $\omega$ gives the basic curve $\gamma$. On the plane $Oyz$
the basic curve $\gamma$ is defined in polar coordinates
$y=r\cos\theta$, $z=r\sin\theta$ by the equation $r=f(\theta)$.
Note, that since the value of $j$ can have arbitrary sign, both
positive and negative values of polar coordinate $r$ are allowed in
the construction of $\gamma$.

The obtained geometrical interpretation provides an algorithm of
construction of the vector field, which is defined by the angle
$\omega$ of deviation from the positive direction of the $Oy$ axis.
Angle $\omega$ is determined from the solutions of implicit equation
(\ref{ImplEqh0}). Suppose, that function $f$ in equation
(\ref{ImplEqh0}) is given. This means, that one can construct the
basic curve $\gamma$ by the formula $r=f(\theta)$ in polar frame of
reference on $Oyz$ plane. Determination of angle $\omega$ at the
point $M=(y,z)$ of the plane $x=x_0$ requires the following
operations as illustrated in figure \ref{fh_0_2}.

\begin{enumerate}
  \item Draw a circle $S_M$ with diameter $OM$.
  \item Find the intersection points $A_i$ of the circle $S_M$ with curve
  $\gamma$. If $S_M$ does not intersect $\gamma$ then $M$ does
  not belong to the domain of $\omega$.
  \item The angle between the line segment $OA_i$
  and a positive direction of $Ox$ axis gives a value of the angle $\omega$
  at point $M$ (see figure \ref{fh_0_2}).
  \item Angle $\omega$ has the same value at all points of the line
  passing through the line segment $A_iM$.
\end{enumerate}

\begin{figure}
\begin{minipage}[b]{0.5\textwidth}
\includegraphics[width=0.8\textwidth]{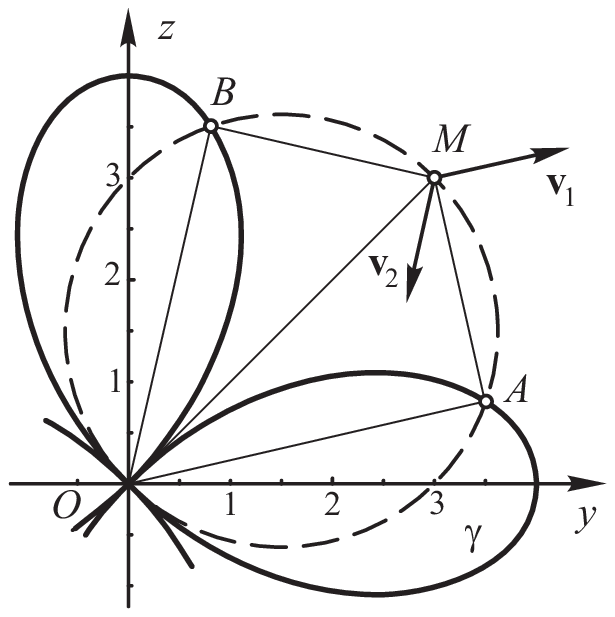}
\caption{Curve a $\gamma$ is defined by equation $r=\cos2\theta$ where
both positive and negative values of $r$ are allowed. Point $B$
corresponds to the part of the curve with negative $r$. The
direction ${\bf v}_2$ assigned to $B$ is therefore opposite to the
one given by the segment $OB$.}\label{fh_0_3}
\end{minipage}
\hspace{3mm}
\begin{minipage}[b]{0.5\textwidth}
\includegraphics[width=0.9\textwidth]{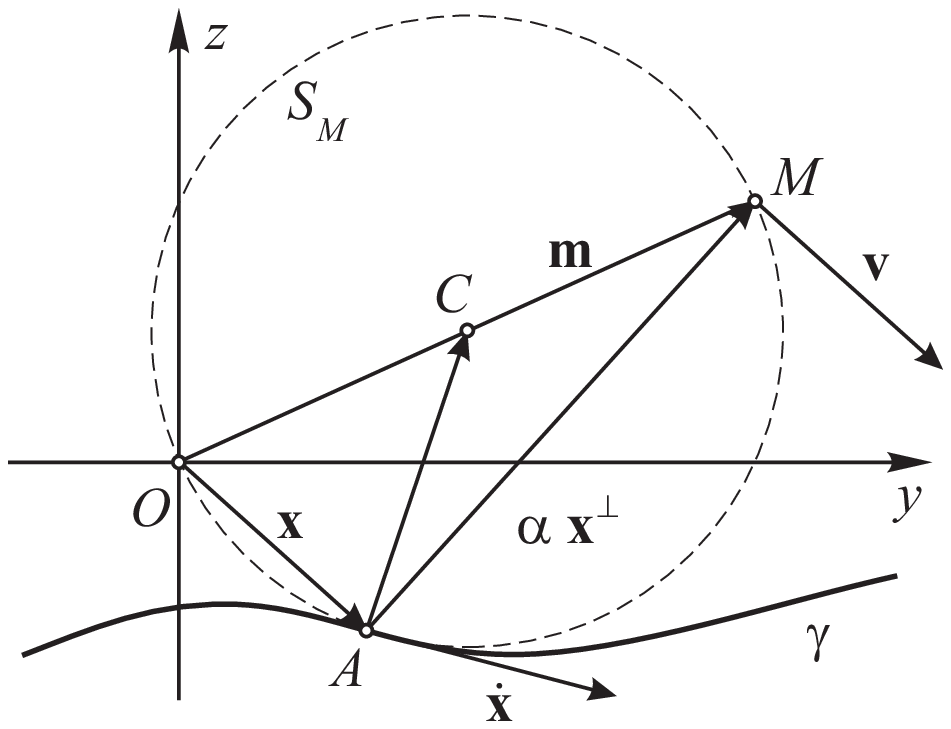}
\caption{Point $M$ belongs to the boundary of the domain of function
$\omega$ if the circle $S_M$ is tangent to $\gamma$ at some point
$A$. From the elementary geometry vectors $OA$ and $AM$ are
orthogonal. This allows expressing vector $\bm$ in terms of $\bx$
and $\bx^\perp$.}\label{fh_0_4}
\end{minipage}
\end{figure}

As mentioned before, function $f$ can be both positive and negative.
Negative $f$ corresponds to negative coordinate $j$. Thus, if the
point of intersection of the curve $\gamma$ and auxiliary circle
$S_M$ belongs to those part of the curve, which corresponds to the
negative values of $f$, then the vector field should be taken with
the negative sign, i.e. instead of $\omega$ one should take
$\omega+\pi$. This situation is shown in figure \ref{fh_0_3}. Here
the curve $\gamma$ is determined by the equation $r=\cos2\theta$.
The circle with diameter $OM$ for $M=(3,3)$ has two points of
intersection with curve $\gamma$. Point $A$ belongs to the
``positive'' part of curve $\gamma$, therefore it defines the
direction $\mathbf{v}_1$, codirectional with the segment $OA$. Point
$B$ lies on the ``negative'' part of $\gamma$, i.e. the
corresponding direction $\mathbf{v}_2$, is opposite to the one,
defined by the segment $OB$.

Next, it is necessary to find the domain of function
$\omega=\omega(t_0,x_0,y,z)$ defined by  the implicit equation
(\ref{ImplEqh0}) over the plane $x=x_0$. Assume that curve $\gamma$
is given. Point $M$ belongs to the boundary of the domain if the
circle $S_M$ with diameter $OM$ is tangent to curve $\gamma$ at some
point $A$ (see figure \ref{fh_0_4}). Let the position vector of
point $M$ be $\bm$. Parametrization of $\gamma$ is taken in the form
$\bx=\bx(s)$ with some parameter $s\in\Delta\subset \mathbb{R}$.
From the elementary geometry $\bm=\bx+\alpha\bx^\perp$, where
$\bx^\perp\cdot\bx=0$. The tangency condition of the circle and
curve $\gamma$ gives $(\bm/2-\bx)\cdot\dot{\bx}=0$. Here and further
the upper dot denotes the differentiation with respect to $s$.
Substitution of the expression for $\bm$ form the first equality
into the second one provides
$(\alpha\bx^\perp/2-\bx/2)\cdot\dot{\bx}=0$. The scalar $\alpha$ is
then determined by
\[\alpha=\frac{\bx\cdot\dot{\bx}}{\bx^\perp\cdot\dot{\bx}}.\]
Thus, the border of the domain of function $\omega$ has the
following parametrization
\begin{equation}\label{domainh0}
\bm=\bx+\frac{\bx\cdot\dot{\bx}}{\bx^\perp\cdot\dot{\bx}}\,\bx^\perp.\;\;\;\bx=\bx(s),\;\;\;
s\in\Delta\subset\mathbb{R}.
\end{equation}
Note, that $\bm$ does not depend on the choice of the sign and
length of $\bx^\perp$. At the border's points the vector field
defined by $\omega$ has $\bx$ direction. This direction is
orthogonal to the border. Indeed,
\[\dot{\bm}\cdot\bx=(\dot{\bx}+\dot{\alpha}\bx^\perp+\alpha\dot{\bx}^\perp)\cdot\bx=
\dot{\bx}\cdot\bx+\frac{\bx\cdot\dot{\bx}}{\bx^\perp\cdot\dot{\bx}}\,\dot{\bx}^\perp\cdot\bx=0.\]
The last expression vanishes because from $\bx\cdot\bx^\perp=0$ it
follows $\dot{\bx}\cdot\bx^\perp=-\bx\cdot\dot{\bx}^\perp.$

\begin{figure} \centering
\includegraphics[width=0.45\textwidth]{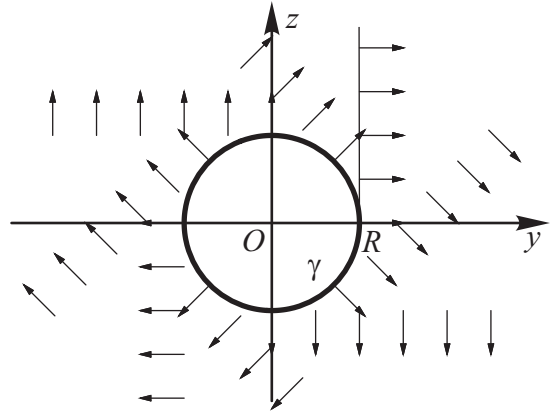}
\caption{The vector field defined by the curve $\gamma:y^2+z^2=R^2$.
}\label{fh_0_5}
\end{figure}
As an example, let us take $\gamma$ to be the circle $y^2+z^2=R^2$.
The border of the domain of $\omega$ in this case coincide with the
circle $\gamma$ because for each point $\bx$ of the border one has
$\bx\cdot\dot{\bx}=0$. The corresponding vector field describes a
flow from the cylindrical source and shown in figure \ref{fh_0_5}.
In limit $R=0$ one obtains a vector field corresponding to the
rotation around the origin.

\section{Particles trajectories and magnetic field lines}
\subsection{Trajectory and magnetic field line pattern}\label{traj}
First of all, let us notice that from equations (\ref{NonInv31}),
(\ref{NonInv32}) for $\rho V^2-N^2\ne0$ follows the equality
\begin{equation}\label{Domega}
D\omega=0.
\end{equation}
The trajectory of each particle is a planar curve. Indeed, equation
(\ref{Domega}) implies that angle $\omega$ has constant value along
each trajectory. Hence, the whole trajectory belongs to the plane,
which is parallel to $Ox$ axis and turned on angle $\omega$ about
this axis. The same holds for a magnetic field line, because
vanishing of the second factor in (\ref{Class3}) is equivalent to
constancy of $\omega$ along each magnetic curve. Thus, for each
particle its trajectory and magnetic field line are planar curves,
which lie in the same plane defined by the angle $\omega$.

\begin{figure}[t]
  \centering
  \includegraphics[width=0.38\textwidth]{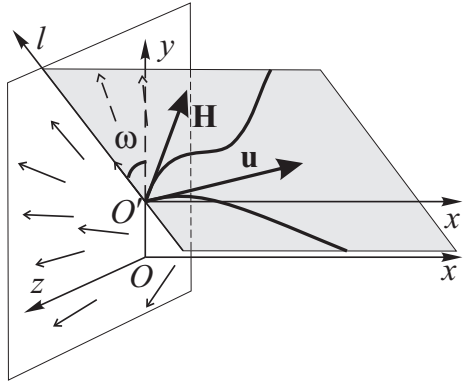}
  \caption{Trajectories and magnetic field lines are planar curves, which are
  the same for all particles, belonging to the same plane
  $x=\const$. In order to determine the flow in the whole space it is required to
  set up an admissible vector field of directions in some plane $x=x_0$ (i.e. to
  determine function $\omega$ consistent with equations
  (\ref{omega3}) or (\ref{ImplEqh0})) and calculate trajectory and
  magnetic field line for arbitrary particle in this plane. The
  whole picture of the flow is obtained by attaching the trajectory and
  the magnetic line pattern to each point on the plane $x=x_0$ in
  accordance with the vector field of directions.
  }\label{newframe}
\end{figure}

The second important property follows from the representation of the
solution (\ref{SolRepr}). Let us set up a Cauchy problems for
trajectory of some particle. The particle moves in its plane, hence
in this plane the motion is completely defined by components of
velocity $U$ and $V$. These two functions depend only on invariant
variables $t$ and $x$. Therefore, for any two particles, which
belong to the same plane $x=x_0$ at initial time $t=t_0$ the Cauchy
problems for the trajectories coincide. Of course, the two different
particles move in their own planes, but both trajectories as planar
curves are exactly the same. Similar observation is true for any two
magnetic lines passing through two different points in the same
plane $x=x_0$. Thus, one can construct a pattern by calculating the
trajectory and the magnetic field line for any particle in the plane
$x=x_0$. The pattern attached to each points in the plane $x=x_0$
inside of the domain of function $\omega$ according to the field of
directions defined by function $\omega$ produces the 3D picture of
trajectories and magnetic field lines in the whole space. The
described algorithm is illustrated in figure \ref{newframe}.

In order to construct the pattern let us observe a plane of motion
of some particle, which is located at initial time $t=t_0$ at some
point $M=(x_0,\,y_0,\,z_0)$. This plane is parallel to $Ox$ axis and
turned about $Ox$ axis on angle $\omega$. Cartesian frame of
reference is defined in the plane of motion as follows. The origin
$O'$ of the frame is placed at the projection of point $M$ into
$Oyz$ plane. One of the coordinate axes is chosen to be parallel to
$Ox$ axis and denotes by the same letter $x$. Another axis $O'l$ is
placed orthogonally to $O'x$ such that the frame $O'xl$ has right
orientation (see figure \ref{newframe}). Particle's trajectory in
this frame of reference is defined by the solution of the Cauchy
problem
\begin{equation}\label{Traject1}
\frac{dx}{dt}=U(t,x),\;\;\;x(t_0)=x_0.
\end{equation}
The dependence $x=x(t,x_0)$ given by a solution of (\ref{Traject1})
allows finding the dependence $l=l(t)$ along the trajectory by the
formula
\begin{equation}\label{Traject2}
l(t)=\int\limits_{t_0}^t V(t,x(t,x_0))dt.
\end{equation}
The planar curve determined by the dependencies $x=x(t,x_0)$ and
$l=l(t)$ forms a pattern of the trajectory for any particle, which
belongs to the plane $x=x_0$ at $t=t_0$. Equations of particle's
trajectory in initial $Oxyz$-frame are restored in the form
\begin{equation}\label{Traject3}
x=x(t,x_0),\;\;\;y=y_0+l(t)\cos\omega_0,\;\;\;z=z_0+l(t)\sin\omega_0.
\end{equation}
Here $\omega_0=\omega(t_0,\bx_0)$ is the value of angle $\omega$
taken at initial time $t=t_0$ at point $M$.

The magnetic field line at $t=t_0$ is an integral curve of the
magnetic vector field. The pattern of the magnetic curve passing at
$t=t_0$ through the plane $x=x_0$ is given by
\[l(x)=\int\limits_{x_0}^x\frac{N(t_0,s)}{H(t_0,s)}ds.\]
Equations of the magnetic field curve in $Oxyz$ frame of reference
are restored as
\begin{equation}\label{MSL}
y=y_0+\cos\omega_0\,\int\limits_{x_0}^x\frac{N(t_0,s)}{H(t_0,s)}ds,\;\;\;
z=z_0+\sin\omega_0\,\int\limits_{x_0}^x\frac{N(t_0,s)}{H(t_0,s)}ds.\;\;\;
\end{equation}
Derivation of these formulae is similar to those given for the
trajectory (\ref{Traject3}).

Thus, the following properties of plasma motion holds (see figure
\ref{newframe}).
\begin{itemize}
\item Trajectories and magnetic lines lie entirely in planes, which are orthogonal
to the $Oyz$-plane and turned on angle $\omega$ about $Ox$ axis.

\item All particles, which belong at some moment of time $t=t_0$ to a
plane $x=x_0$, circumscribe the same trajectories in planes of each
particle motion. Magnetic field lines passing through a plane
$x=x_0$ are also the same planar curves.

\item Angle of rotation about $Ox$-axis of the plane containing the trajectory and the magnetic line
of each particle is given by function $\omega$, which satisfies
equation (\ref{omega3}) or (\ref{ImplEqh0}).
\end{itemize}

\subsection{Domain of the solution in 3D space}

The constructions above show that the whole area in 3D space
occupied by the solution is obtained as follows. In fixed plane
$x=x_0$ function $\omega$ has some (in many cases, finite)
definition domain, bounded by $\tau$-equidistants to $\gamma$ for
$h\ne0$ and by the curve \eqref{domainh0} for $h=0$. In both cases
the field of direction defined by $\omega$ in $x=x_0$ plane is
orthogonal to the boundary of the $\omega$-domain. In order to
obtain boundaries of the whole 3D domain of the solution one should
attach the magnetic line pattern, calculated on some particular
solution of the invariant system, to every point of the boundary of
$\omega$-domain in plane $x=x_0$ according to the usual algorithm.
This gives a canal woven from magnetic lines which pass through
boundaries of the $\omega(t_,x_0,y,z)$-domain and intersect $x=x_0$
plane. The walls of the canal can be interpreted as rigid infinitely
conducting pistons. Due to the well-known property of magnetic field
line freezing-in, the walls are impermeable for plasma. In case of
stationary solution the walls are fixed. In non-stationary case the
walls extend or shrink according to the behavior of function $\tau$
for $h\ne0$ and $\varphi$ for $h=0$. In case of finite
$\omega$-domain (it can always be restricted to a finite one) each
$x$-cross-section of the 3D-domain of the solution is finite,
therefore both magnetic and kinetic energy have finite value in each
$x$-layer.

\subsection{Stationary flow}

As an example we observe a stationary solution of system
\eqref{main1}--\eqref{main6}. Suppose that all unknown functions
depend on $x$ only. This leads to the following system of ODEs:
\begin{eqnarray}\label{stst1}
&&U\rho'+\rho(U'+hV)=0.\\\label{stst2}
&&UU'+\rho^{-1}p'+\rho^{-1}NN'=0.\\\label{stst3}
&&UV'-\rho^{-1}H_0hN'=0,\\\label{stst4}
&&Up'+A(p,\rho)(U'+hV)=0,\\\label{stst5}
&&UN'+NU'-H_0hV'+hNV=0,\\\label{stst6}
&&Uh'+Vh^2=0,\;\;\;H_0h'+hN=0.
\end{eqnarray}
Elimination of the derivative $h'$ in equations \eqref{stst6} gives
the finite relation
\begin{equation}\label{Int31}
H_0Vh=UN,
\end{equation}
which states collinearity of the magnetic and velocity fields at
each particle. The same property holds for the analogous spherical
solution \cite{GolovinSingVortMHDInvSubm}. Equation \eqref{stst5} is
satisfied identically by virtue of \eqref{Int31}.

Equation \eqref{stst4} gives entropy conservation
\begin{equation}\label{Int32}
S=S_0.
\end{equation}
Equation \eqref{stst1} under condition \eqref{Int31} gives the flow
rate integral
\begin{equation}\label{Int33}
\rho\, U=nh,\;\;\;n=\const.
\end{equation}
Substitution of the obtained integrals into \eqref{stst3} allows
finding the following relation between the tangential components of
velocity and magnetic fields
\begin{equation}\label{Int34}
nV-H_0N=m,\;\;\;m=\const.
\end{equation}
Integration of equation \eqref{stst2} gives the Bernoulli integral
\begin{equation}\label{Int35}
U^2+V^2+2\int \frac{dp}{\rho}=b^2,\;\;\;b=\const.
\end{equation}

The only equation left to integrate is any of two equations
\eqref{stst6}. With its aid all unknown functions may be expressed
in terms of the "potential" $\tau=1/h$ as
\begin{equation}\label{mn115}
U=\frac{m\tau+H_0^2\tau'}{n\tau\tau'},\;\;\;V=\frac{m\tau+H_0^2\tau'}{n\tau},\;\;\;H=\frac{H_0}{\tau},
\;\;\;N=\frac{H_0\tau'}{\tau},\;\;\;\rho=\frac{n^2\tau'}{m\tau+H_0^2\tau'}.
\end{equation}

a) Let $m\ne 0$. Using the admissible dilatations it is convenient
to make $m=n=\mathrm{sign}(\tau\tau')$. Expressions \eqref{mn115}
become
\begin{equation}\label{mn11}
U=\frac{\tau+H_0^2\tau'}{\tau\tau'},\;\;\;V=\frac{\tau+H_0^2\tau'}{\tau},\;\;\;H=\frac{H_0}{\tau},
\;\;\;N=\frac{H_0\tau'}{\tau},\;\;\;\rho=\frac{\tau'}{\tau+H_0^2\tau'}.
\end{equation}
Substitution of \eqref{mn11} into the Bernoulli integral
\eqref{Int35} produces an equation for $\tau$. In case of polytropic
gas with the state equation $p=S\rho^\gamma$ it has the following
form
\begin{equation}\label{KeyEqStat}
\left(\frac{\tau+H_0^2\tau'}{\tau\tau'}\right)^2+\left(\frac{\tau+H_0^2\tau'}{\tau}\right)^2+
\frac{2\gamma
S_0}{\gamma-1}\left(\frac{\tau'}{\tau+H_0^2\tau'}\right)^{\gamma-1}=b^2.
\end{equation}
This ODE for $\tau(x)$ is not resolved with respect to the
derivative $\tau'$, which complicates its investigation. Examples of
analysis of such non-resolved ODEs can be found in
\cite{Chup1}--\cite{Pavl}. One can show that there are several
branches of solution $\tau(x)$ of equation \eqref{KeyEqStat} passing
through each point in $(x,\tau)$ plane, which correspond to
different relations between the velocity $U$ and the characteristics
speeds of MHD system \eqref{MHD-cont}--\eqref{MHD-Faradey}. It is
possible to switch between different branches of the solution via
fast or slow shock waves. However, this investigation lies outside
of the scope of this paper.

b) In case $m=0$ after some straightforward simplifications we
obtain the following solution of system
(\ref{main1})--(\ref{main6}):
\begin{equation}\label{SolStstSpec}
\begin{array}{l}
U=H_0^2\,\mathrm{sech}\, x,\;\;V=H_0^2\tanh x,\;\;\tau=\cosh x,\\[5mm]
H=H_0\,\mathrm{sech}\, x,\;\;N=H_0\tanh x,\;\;
\rho=H_0^{-2},\;\;\;S=S_0.
\end{array}
\end{equation}
One can check that (\ref{SolStstSpec}) represents a special case of
the more general S. Chandrasekhar solution \cite{Chandra}. This
solution is also invariant with respect to infinite group of
Bogoyavlenskij transformations \cite{Bogoyavlenskij}. The simplicity
of solution \eqref{SolStstSpec} gives opportunity to use it for
demonstration of geometrical algorithms given in previous sections.

Streamlines and magnetic field lines coincide and are given by
formulas (\ref{Traject3}) with $x_0=0$ and
\begin{equation}\label{TrajectStatSpec}
l(x)=\cosh  x-1.
\end{equation}
In each plane of particle's motion the streamline is a half of
catenary. Note, that solution (\ref{SolStstSpec}) can be
continuously adjoined with the uniform flow along $Ox$ axis. Indeed,
in section $x=0$ all functions in (\ref{SolStstSpec}) and their
derivatives take values compatible with the uniform flow. Let us
construct a solution, which switches the uniform flow to the
generalized one-dimensional solution (\ref{SolStstSpec}) at the
section $x=0$. The corresponding streamline is a straight lines for
$x<0$ and a half of the catenary for $x\ge0$. In order to get the
whole three-dimensional picture of motion this streamline pattern
should be attached to each point of the plane $x=0$ according to the
direction field defined by function $\omega$.

\begin{figure}[t]
  \centering
  \includegraphics[width=\the\textwidth]{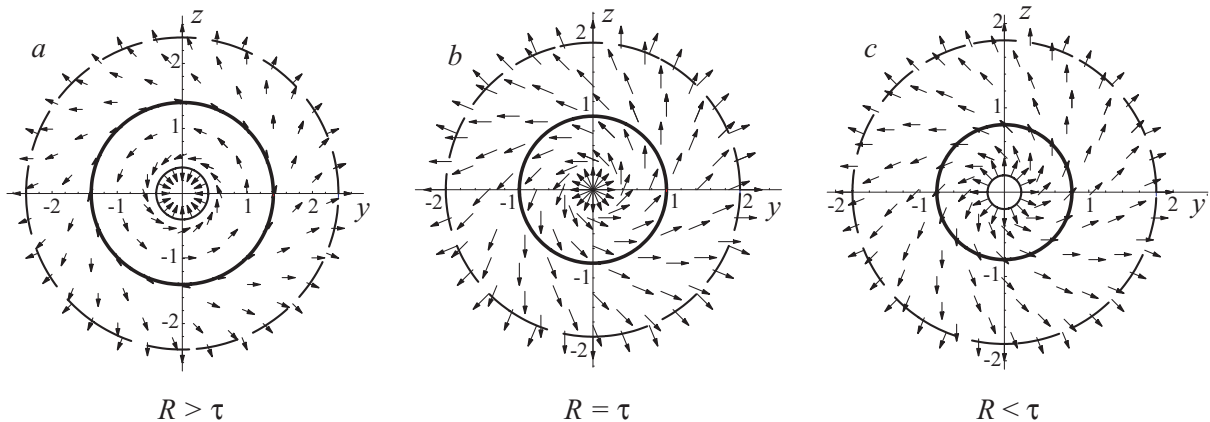}
  \caption{Field of direction obtained by the algorithm of section
  \ref{smain}. Here $\gamma$ is the middle circle of radius $R$.
  Three cases according to the relation between $R$ and $\tau$ distinguishes.
  In all cases the domain of the vector fields is an annular stripe of determinacy
  between two equidistant curves (inner and outer circles in the diagrams).}\label{VectorCircle}
\end{figure}

Function $\omega$ is determined by the implicit equation
(\ref{omega3}). Algorithm of section \ref{smain} requires assigning
some particular function $F$, or some curve $\gamma:\,F(y,z)=0$. Let
the curve $\gamma$ be a circle $y^2+z^2=R^2$. The corresponding
function $\omega$ is determined at each point of the plane $x=0$ by
equation (\ref{omega3}). Figure \ref{VectorCircle} shows the vector
fields obtained for different relations between $\tau$ and $R$. For
$R>\tau$ the vector field is defined in the annular area between two
circles of radii $R\pm\tau$. On the inner equidistant circle
$|\bx|=R-\tau$ the vector field is directed outside of the stripe of
determinacy towards the origin. In case $R=\tau$ the inner
equidistant circle shrinks into the origin $\bx=0$. At that, the
vector field becomes multiply-determined at this point. Finally, for
$R<\tau$ the inner equidistant turns inside out and becomes a circle
of radius $\tau-R$ with the vector field on it directed inside of
the stripe of the determinacy. These three vector fields generate
different pictures of motion in whole 3D space.
\begin{figure}[t]
  \centering
  \includegraphics[width=0.8\textwidth]{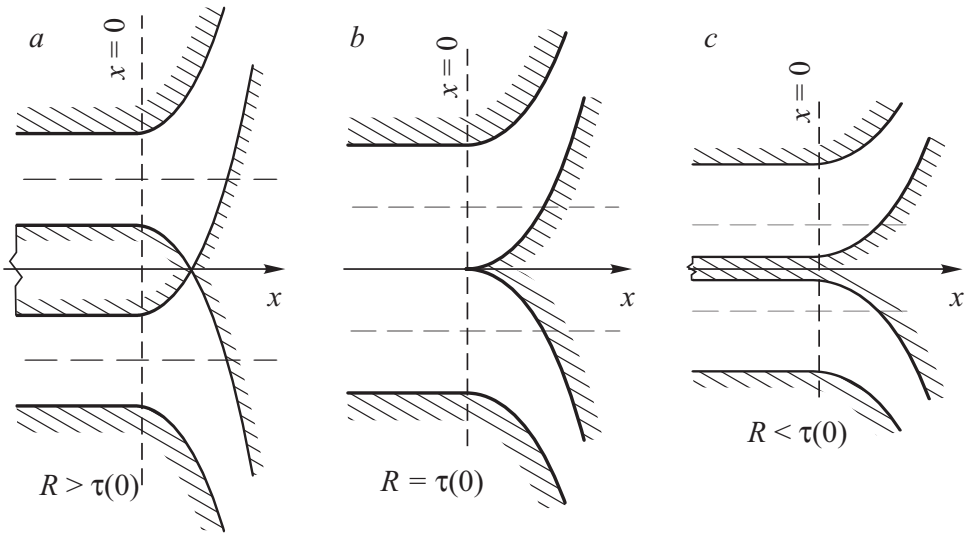}
\caption{Axial sections of axially-symmetrical canal occupied by the
plasma flows. The uniform flow in cylindrical canal for $x<0$
switches at section $x=0$ to the flow in the curvilinear canal for
$x>0$ described by the solution (\ref{SolStstSpec}). The boundary of
the canal is a rigid wall. Cases {\it a}, {\it b} and {\it c}
correspond to the vector fields in figure \ref{VectorCircle}. In the
diagrams {\it a} and {\it c} the canal has an inner cylindrical
core.}\label{Canal}
\end{figure}

The streamline pattern described above should be attached to each
points of $Oyz$ plane inside the corresponding domain of $\omega$
according to the directional fields shown in figure
\ref{VectorCircle}. Because of the obvious central symmetry of the
vector fields the whole picture of motion is axially-symmetrical.
The axial section of the area in 3D space, occupied by the
corresponding flow is shown in figure \ref{Canal}.

\begin{figure}[t]
\begin{center}
  \includegraphics[width=0.6\textwidth]{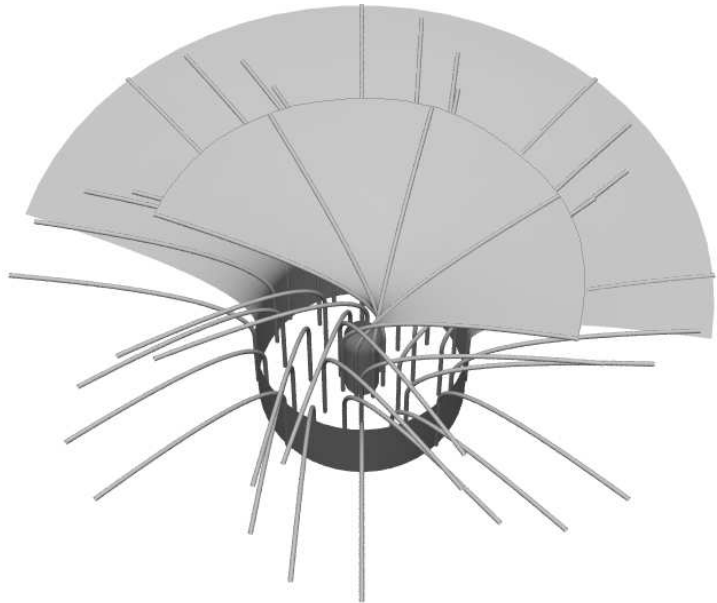}
  \caption{Tree-dimensional visualization of motion.
  Fragments of the canal's walls and the streamlines are shown. Each
  streamline has a shape of the same flat curve. Orientation of each
  streamline is defined by the vector field in figure
  \ref{VectorCircle}{\it a}. The axial section of the canal
  is represented in figure \ref{Canal}{\it a}.
  }\label{3Dviz}
  \end{center}
\end{figure}

We assume that uniform flow for $x<0$ changes at section $x=0$ to
the flow, described by the solution (\ref{SolStstSpec}). Depending
on the relation between $\tau(0)$ and $R$ three different pictures
of motion are possible. Each particle moves along the same planar
curve, however orientation of the streamlines in the space differ
from one particle to another. Three-dimensional visualization of the
motion for $R>\tau(0)$ is shown in figure \ref{3Dviz}.

\section*{Conclusion} In present work a new solution of ideal fluid
dynamics equations, describing three-dimen\-sional motions of
plasma, gas and liquid is constructed. The solution is determined by
a system of equations with two independent variables, which is
analogous to the classical system for one-dimensional fluid motions.
At that, the new solution describes spatial nonlinear processes and
singularities, which are impossible to obtain in the classical
framework.

In the constructed solution particles trajectories and magnetic
field lines are flat curves. Trajectory of each curve and its
magnetic field line belong to the same plane parallel to $Ox$ axis.
In contrast to the classical one-dimensional solution, plane of
motion of each particle has its own orientation, which is given by
an additional finite relation. The functional arbitrariness of the
finite relation allows varying the geometry of obtained motion in
accordance to the problem under consideration. Depending on the
chosen geometry, singularities on the border of the region, occupied
by fluid, may appear. In such cases particles may collide at the
border of the domain of the flow. The criterion of singularities
appearance in terms of invariant properties of the arbitrary
function, which specifies the geometry of the flow is given.

The obtained solution may be used as a test for numerical modeling
of complicated three-dimensional flows of infinitely conducting
plasma. It also may serve for theoretical investigations of
three-dimensional singularities of the ideal fluid and plasma
motions.

\section*{Acknowledgements}
Author would like to thank professor O.I. Bogoyavlenskij and
Mathematical \& Statistical Department of Queen's University for
hospitality and inspiring atmosphere, which stimulated this
research. Author also acknowledge the support of Russian Foundation
for Basic Research (project 05-01-00080), of President Programme of
Support of the Leading Scientific Schools (grant
Sc.Sch.-5245.2006.1), and of Integration Project 2.15 of Siberian
Branch of RAS.

\end{document}